\documentclass[twocolumn,english]{revtex4-1}
\usepackage[T1]{fontenc}
\usepackage[latin9]{inputenc}
\usepackage{graphicx}

\makeatletter

\providecommand{\tabularnewline}{\\}

\usepackage{babel}

\usepackage{babel}

\makeatother

\usepackage{babel}
\begin{document}

\title{Electron spin resonance study of atomic hydrogen stabilized in solid
neon below 1$\,$K.}

\author{S.$\,$Sheludiakov}

\altaffiliation{Present address: Institute for Quantum Science and Engineering, Department of Physics and Astronomy, Texas A\&M University, College Station, TX, 77843, USA}

\selectlanguage{english}%

\affiliation{Department of Physics and Astronomy, University of Turku, 20014 Turku,
Finland }

\author{J.$\,$Ahokas}

\affiliation{Department of Physics and Astronomy, University of Turku, 20014 Turku,
Finland }

\author{J.$\,$Järvinen}

\affiliation{Department of Physics and Astronomy, University of Turku, 20014 Turku,
Finland }

\author{L.$\,$Lehtonen}

\affiliation{Department of Physics and Astronomy, University of Turku, 20014 Turku,
Finland }

\author{S.$\,$Vasiliev}

\email{servas@utu.fi}

\selectlanguage{english}%

\affiliation{Department of Physics and Astronomy, University of Turku, 20014 Turku,
Finland }

\author{Yu.$\,$A.$\,$Dmitriev}

\affiliation{Ioffe Institute RAS, 26 Politekhnicheskaya,
St. Petersburg 194021, Russian Federation }

\author{D.$\,$M.$\,$Lee}

\affiliation{Institute for Quantum Science and Engineering, Department of Physics
and Astronomy, Texas A\&M University, College Station, TX, 77843,
USA}

\author{V.$\,$V.$\,$Khmelenko}

\affiliation{Institute for Quantum Science and Engineering, Department of Physics
and Astronomy, Texas A\&M University, College Station, TX, 77843,
USA}

\date{\today}
\begin{abstract}
We report on an electron spin resonance study of atomic hydrogen stabilized
in a solid Ne matrices carried out at a high field of 4.6$\,$T and
temperatures below 1$\,$K. The films of Ne, slowly deposited on the
substrate at the temperature $\sim$1$\,$K, exhibited a high degree
of porosity. We found that H atoms may be trapped in two different
substitutional positions in the Ne lattice as well as inside clusters
of pure molecular H$_{2}$ in the pores of the Ne film. The latter
type of atoms was very unstable against recombination at temperatures
0.3-0.6$\,$K. Based on the observed nearly instant decays after rapid
small increase of temperature, we evaluate the lower limit of the
recombination rate constant $k_{r}\geq5\cdot10^{-20}\,$cm$^{3}$s$^{-1}$
at 0.6$\,$K, five orders of magnitude larger than that previously
found in the thin films of pure H$_{2}$ at the same temperature.
Such behavior assumes a very high mobility of atoms and may indicate
a solid-to-liquid transition for H$_{2}$ clusters of certain sizes,
similar to that observed in experiments with H$_{2}$ clusters inside
helium droplets (Phys. Rev. Lett 101, 205301 (2008)). We found that
the efficiency of dissociation of H$_{2}$ in neon films is enhanced
by 2 orders of magnitude compared to that in pure H$_{2}$, which
is instigated by a strong emission of secondary electrons. 
\end{abstract}
\maketitle

\section{Introduction}

Quantum properties of insulating solids are most pronounced for molecular
and atomic crystals of the lightest elements H$_{2}$, He, Ne. While
making a solid from helium requires applying high pressure even at
zero temperature, hydrogen and neon solidify at ambient pressure and
fairly high temperatures. These solids may serve as matrices for stabilizing
unstable and highly reactive species and radicals. Being introduced
into an inert solid matrix, these species become immobilized and remain
stable at low enough temperatures where their diffusion is suppressed.
Matrices made out of mixtures of neon and hydrogen are most intriguing
because the Lennard-Jones potential parameters for the H$_{2}$-H$_{2}$
and H$_{2}$-Ne pairwise interaction are nearly identical \cite{Gal'tsov04}
while a large difference of their masses allows considering H$_{2}$
and Ne as isotopes of the same substance with an astonishingly different
degree of quantumness.

The solid solutions of H$_{2}$-Ne may form various interesting phases.
It turns out that the equilibrium solubility of H$_{2}$ in solid
neon is vanishingly small and does not exceed a fraction of a percent
for films crystallized from liquids. However, the situation is different
for the non-equilibrium samples prepared by rapid flash-condensing
of the films. Such non-equilibrium samples may contain a co-existing
metastable $hcp$ phase of solid Ne along with the $fcc$ and $hcp$
structures of Ne and H$_{2}$, respectively \cite{Gal'tsov04}. The
phases of molecular hydrogen appear in the form of nanocluster inside
solid neon already at concentrations as low as 0.01$\%$ \cite{Zhitnikov87}.
It has also been observed that solid rare-gas films quench-condensed
on a cold substrate were found to be highly disordered and porous
\cite{Schulze74}.

The possibility of having small hydrogen clusters in neon crystals
has another important consequence, since it has been predicted that
in a restricted geometry the freezing temperature may be substantially
lowered and one may supercool liquid hydrogen to the superfluid transition,
which may occur at a temperature of $\sim$6$\,$K \cite{Ginzburg72}.
The liquid-like behavior of hydrogen clusters surrounded by superfluid
helium film has been reported for clusters composed of $\sim$10000
molecules at temperature of $\sim$1$\,$K \cite{Kuyanov08}, and
for small clusters of less than 30 molecules, a superfluid response
was predicted \cite{Sindzingre91,Kwon02,Mezzacapo06} and observed
at 0.15$\,$K \cite{Grebenev00}. This observation, however, was somewhat
controversial \cite{Callegari01,Omiyinka14} resulting in an uncertainty
for the behavior of clusters of intermediate size.

Hydrogen atoms can be obtained inside solid matrices by dissociation
of molecular H$_{2}$ using e.g. electrons or $\gamma$-rays. Due
to their small size, H atoms may occupy different lattice positions
in the matrix where their interaction with the host particles may
slightly change both their electronic $g$-factors and the hyperfine
constants as compared with those of free atoms in the gas phase \cite{Adrian60}.
The weak interatomic van-der-Waals attractive interaction, Ne-Ne,
can be easily overcome if the H atom is introduced into an interstitial
position of the neon matrix. In this case, a strong Ne-H repulsion
which appears at short distances rearranges the host atoms around
the hydrogen guests in a way such that the H atoms finally reside
in the substitutional positions. This process, known as relaxation
\cite{Hall57}, also takes place in the matrices of solid hydrogen
isotopes where a strong H-H$_{2}$ repulsion does not allow stabilizing
H atoms in the interstitial sites \cite{Li94}. Relaxation of H atoms
in the rare-gas solids was considered by Kiljunen et al. \cite{Kiljunen99}
who calculated that H atoms in solid Ne will reside in the substitutional
positions regardless of their initial trapping sites.

Exposing solidified rare gases to an electron beam results in large
yields of secondary electrons, as well as atomic and molecular excitons
\cite{Schou86}. This is in stark difference to molecular solids,
such as H$_{2}$ and D$_{2}$, where primary electrons can lose their
energy by exciting rotational and vibrational transitions or by dissociating
molecules \cite{Schou78}. The large quantity of secondary electrons
is very effective for increasing the dissociation efficiency of H$_{2}$
molecules embedded in rare-gas matrices.

Stabilization of hydrogen atoms in solid neon has a long and controversial
history. The first attempts to stabilize H atoms by co-deposition
of the rf discharge products onto a cold substrate appeared to be
unsuccessful, where only a single ESR line doublet was observed after\emph{
in situ} photolysis of HI \cite{Foner60}. This ESR line doublet was
characterized by a positive hyperfine constant change, $\frac{\Delta A}{A}=0.43\%$,
and was attributed to the H atoms in the somewhat distorted substitutional
positions of the Ne matrix. Later Zhitnikov and Dmitriev used a co-deposition
technique and reported the observation of an extremely narrow, $\sim$80$\,$mG
wide, ESR line doublet with a negative hyperfine constant change $\frac{\Delta A}{A}\mbox{=}-0.1\%$
\cite{Zhitnikov87} which they attributed to H atoms in substitutional
sites of an unperturbed solid Ne lattice. In a following work \cite{Dmitriev89},
Dmitriev et al. reported on the observation of two positions for H
atoms in solid Ne characterized by both positive and negative hyperfine
constant changes respectively. Along with a single positively shifted
ESR line doublet of isolated atoms in solid Ne, Knight et al. and
Correnti et. al. observed the ESR spectra of H-H and H-D radical pairs
\cite{Knight98} and also H$_{2}^{+}$ ions \cite{Correnti12} unstable
in a hydrogen environment, respectively, using the same co-deposition
technique.

In this work, we report on the first ESR study of atomic hydrogen
trapped in solid Ne carried out in a high magnetic field (4.6$\,$T)
and at temperatures below 1$\,$K. We found that the as-deposited
Ne samples are highly disordered and porous, although the porosity
can be significantly reduced by annealing the films at 7-10$\,$K.
The H$_{2}$ molecules in solid Ne were dissociated \emph{in situ}
by electrons released during the $\beta$-decay of tritium trapped
in the metal walls of our sample cell. The accumulation rate of H
atoms from dissociation of H$_{2}$ molecules in solid neon turned
out to be enhanced by two orders of magnitude as compared with that
in pure H$_{2}$. We attribute this to dissociation of H$_{2}$ molecules
by secondary electrons released from neon atoms during the course
of their bombardment by primary electrons generated in the $\beta$-decay
of tritium. The ESR lines of H atoms in the as-deposited solid neon
films had a complex shape which can be fitted by a sum of three peaks.
These components were assigned to H atoms in different positions in
the matrix, two of which correspond to the substitutional positions
in the Ne lattice, while the third originates from the H atoms trapped
in the regions of pure H$_{2}$. We found that raising the sample
cell temperature from 0.1$\,$K to 0.3-0.6$\,$K leads to an abrupt
recombination of the atoms inside regions of pure H$_{2}$. Such rapid
recombination cannot occur at 0.6$\,$K inside a solid. We will consider
several possibilities for the explanation of this effect.

\section{Experimental details}

\begin{figure}
\includegraphics[width=1\columnwidth]{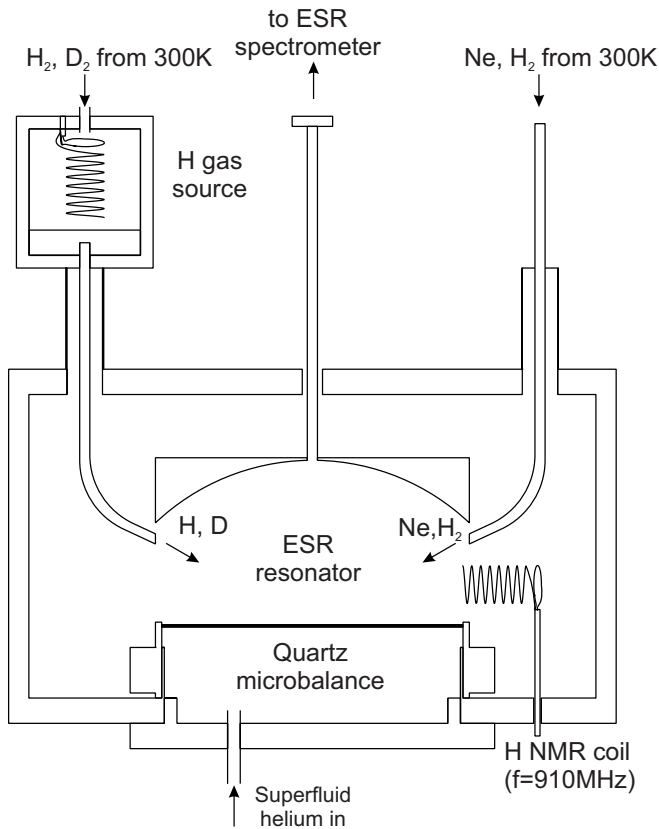}

\protect\caption{Sample cell schematic. The auxiliary rf resonator used for ENDOR is
labeled as H NMR coil. \label{fig:Cell}}
\end{figure}

The experiments were carried out in the sample cell (SC) shown in
Fig.$\,$\ref{fig:Cell} \cite{Cellpaper}. The cell is attached to
the mixing chamber of an Oxford 2000 dilution refrigerator to provide
cooling. The minimum temperature attained in this set of experiments
was $\simeq$90$\,$mK. The solid neon samples were deposited directly
from a room temperature gas handling system at a deposition rate of
1-2 monolayers/s. We used a 99.99$\,$\% pure neon gas and did not
add H$_{2}$ intentionally while we prepared so-called ``pure''
Ne samples. The mass-spectrometry analysis of the neon gas we used
showed a H$_{2}$ content $\approx$100$\,$ppm. The samples were
deposited onto the top electrode of a quartz-crystal microbalance
(QM) which has a mass resolution of about 0.02 Ne monolayer. The sample
cell temperature during the film deposition was stabilized at 0.8-1.3$\,$K.
Prior to deposition, a small ($\sim$mmol) amount of helium was condensed
into a volume under the QM (see Fig.$\,$\ref{fig:Cell}) for removing
heat released during the film growth. The top electrode of the quartz
microbalance is electrically insulated from the sample cell body which
made it possible to apply electric potentials to the film substrate.

\begin{figure}
\includegraphics[width=1\columnwidth]{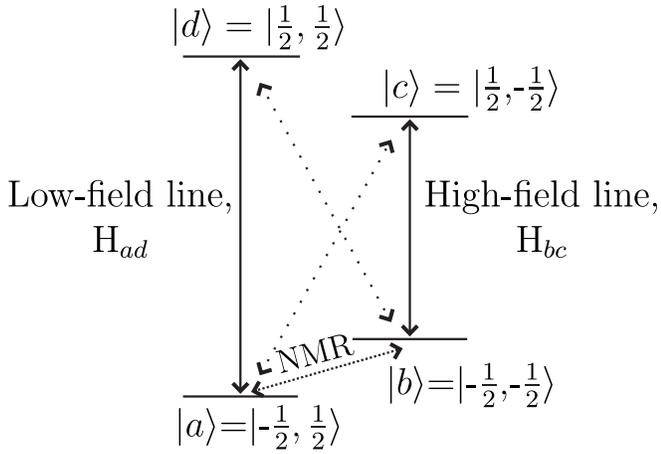}

\protect\caption{Energy level diagram of H atom in magnetic field. Solid arrows mark
the allowed ESR transitions, dotted arrows - the forbidden transitions
corresponding to a simultaneous spin flip of electron ($S$) and proton
($I$). The spin states are labeled as $|m_{S},m_{I}\rangle$. \label{fig:LevelDiag1}}
\end{figure}

The QM top electrode also serves as a flat mirror of the ESR Fabry-Perot
resonator. This allows a simultaneous measurement of the film thickness
and ESR detection of the species in the film which possess unpaired
electron spins. The main investigation tool in our work is a 128$\,$GHz
super-heterodyne ESR spectrometer which enables a simultaneous measurement
of the real and imaginary components of rf magnetic susceptibility
\cite{Vasilyev04}. We will present only the ESR absorption spectra
throughout the article. The spectrometer has a sensitivity of about
10$^{11}$ spins at the excitation power of the order of 1$\,$pW,
while the maximum ESR excitation power available is of the order of
1 $\mu$W. The measurement of the H atom hyperfine constants was carried
out by the method of electron-nuclear double resonance (ENDOR) which
allows determination of the NMR transition frequency indirectly, by
its influence on the ESR signal amplitude \cite{Feher56}. An auxiliary
rf resonator (H NMR coil in Fig.$\,$\ref{fig:Cell}) for performing
ENDOR was arranged close to the top electrode of the quartz microbalance.
The resonator was used to excite the $a-b$ NMR transition of hydrogen
atoms ($f=910\,$MHz) (Fig.$\,$\ref{fig:LevelDiag1}).

Construction of the sample cell also allows accumulation of H and
D atoms in the gas phase at low densities. Recording ESR spectra from
these atoms provides a reference for measurements of the ESR spectrum
parameters of matrix isolated atoms by measuring the shifts of the
ESR lines from positions of the lines of the gas-phase atoms. For
filling the sample cell with hydrogen gas we use a cryogenic source
of atoms (H-gas source) \cite{Disso,Helffrich87} (a separate chamber
located $\sim10$ cm above the sample cell). Small amounts of molecular
hydrogen are condensed into the H-gas source, and then the molecules
are dissociated by running the RF discharge in a miniature coil inside
the source. In order to suppress surface adsorption and recombination
of atoms, the walls of the chamber are covered by a film of superfluid
helium.

\section{Experimental results}

This work follows our previous study of unpaired atoms of hydrogen
isotopes in solid films containing tritium \cite{Tritium_paper}.
Electrons with average energy of 5.7$\,$keV resulting from tritium
$\beta$-decay serve as an effective source of atoms due to \emph{in
situ} dissociation of molecules in the matrix. It turned out that
in the course of previous experiments, substantial numbers of tritium
atoms and molecules became trapped in the ESR resonator mirrors and
possibly in the copper walls of the sample cell. We found that even
warming the sample cell to room temperature and pumping it to high
vacuum for several days does not completely remove tritium from the
walls of the chamber. From the efficiency of production of stabilized
H atoms we estimate that $\sim10^{15}$ tritium atoms still remained
trapped in the walls of the sample cell after cooling the sample cell
to low temperatures.

Our original idea for studying neon was to cover the flat mirror of
the ESR resonator by an inert absorptive layer which would capture
electrons resulting from the decay of trapped tritium. Already in
the first experiment with so-called ``pure'' (99.99\%) solid neon
films, we noticed an unexpectedly high accumulation rate of H atoms,
$d[\mbox{H}]/dt$, resulting from dissociation of trace amounts of
H$_{2}$ molecules condensed with the neon gas. The ESR signal from
H atoms in this neon film was only 20 times smaller than for H atoms
trapped in a pure H$_{2}$ film of the same thickness. Followed by
that, we carried out a series of experiments with solid neon-hydrogen
samples with different admixtures of H$_{2}$, both as-deposited and
annealed. We studied the following samples: a ``pure'' Ne film (Sample
1), Ne:0.2\% H$_{2}$ (Sample 2), Ne:1\% H$_{2}$ (Sample 3), Ne:3\%
H$_{2}$ (Sample 4), Ne:6\% H$_{2}$ (Sample 5). Sample 6 was a ``pure''
Ne film, first studied as-deposited and then annealed and later a
160$\,$nm pure H$_{2}$ film was deposited on top of it. All neon
films had a thickness of 2.5$\,\mu$m. We also studied a pure normal
H$_{2}$ sample of the same thickness for the sake of comparison (Sample
7).

A typical experimental cycle included the following stages: deposition
of the sample at the sample cell temperature 0.7-1.3$\,$K, cooling
down to the lowest temperature $\approx$0.1$\,$K, waiting for 1-3
days for accumulation of the atoms in the deposited film. After that
we usually condensed a small amount of $^{4}$He into the sample cell
in order to accumulate atoms in the gas phase and make accurate measurement
of the spectroscopic parameters of H atoms in the matrix with respect
to that of the free atoms. It turned out that admission of helium
influenced properties of the samples substantially, which could be
related to the overheating of the sample cell to 0.6$\,$K during
condensation. Therefore, for some samples prior to the film condensing,
we performed a brief test of the sample reaction during the increase
of temperature to 0.6$\,$K.

It turned out that all as-deposited samples absorbed a substantial
amount of helium, which indicated the high porosity of the films.
In order to eliminate this, we performed annealing of the films by
heating them to the temperatures of 7-9$\,$K for 1-2 hours. This
procedure was performed close to the end of the experimental cycle
before destroying the sample.

ESR spectra of all visible components within $\pm$500$\,$G from
the H doublet center were periodically recorded at all stages of the
experimental cycle. ENDOR studies were usually performed after accumulation
of the H atoms to strong enough ESR signal strength.

\subsection{ESR spectra}

\begin{figure}
\includegraphics[width=1\columnwidth]{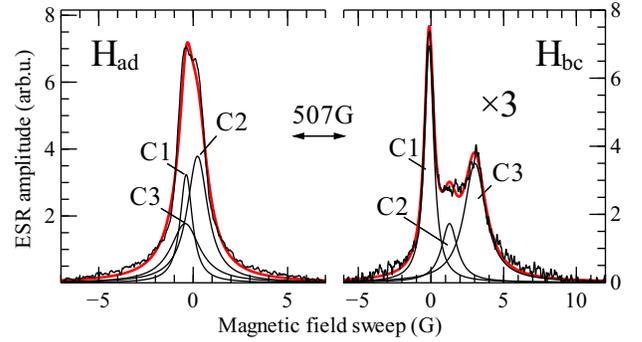}\protect\caption{ESR spectra of atomic hydrogen in Sample 3 (Ne:1\%H$_{2}$) after
2 days of storage. The H$_{bc}$ and H$_{ad}$ lines are fitted by
three Lorentzian curves with parameters shown in Table \ref{tab:spectroscopic-parameters}.
The H$_{bc}$ line amplitude is multiplied by a factor of 3.\label{fig:ESR-spectra-of}}
\end{figure}

\begin{figure}
\includegraphics[width=1\columnwidth]{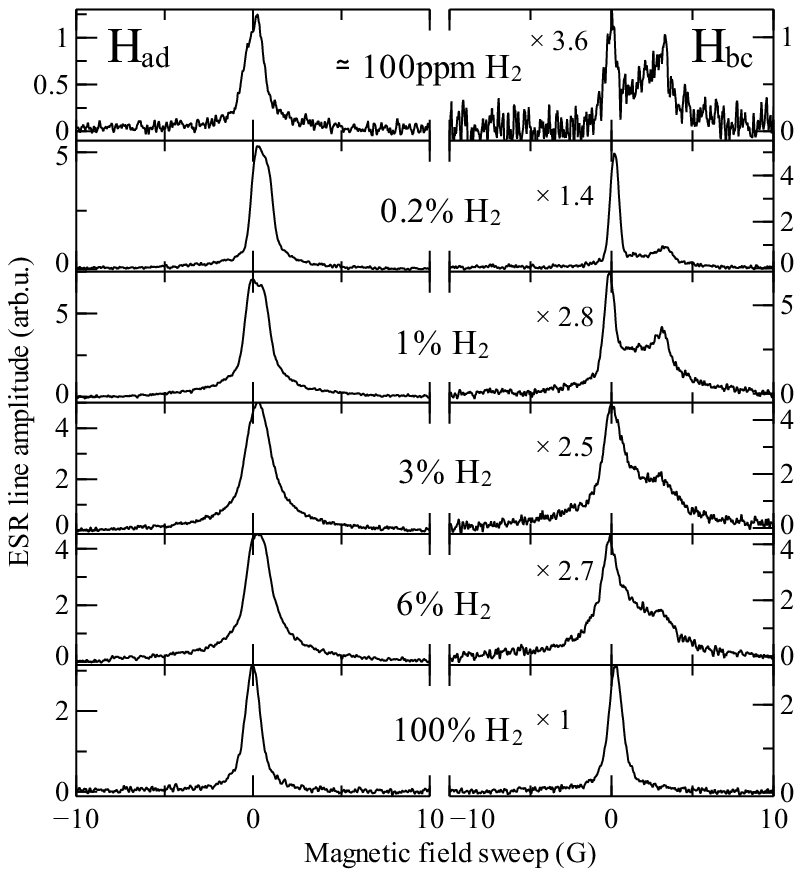}\protect\caption{The H$_{ad}$ and H$_{bc}$ ESR lines measured in neon samples with
different admixtures of H$_{2}$ studied in this work. Note that the
H$_{bc}$ line amplitudes are multiplied by a factor specified for
each plot. The distance between the lines is $\simeq$507$\,$G. \label{fig:The-ESR-spectra}}
\end{figure}

The high-field H$_{bc}$ ESR lines of the H doublet in all as-deposited
samples had a complex shape and could be fitted by 3 Lorentzian lines
shifted from each other. In contrast, the H$_{ad}$ lines had a much
more regular, nearly Lorentzian shape (Figs. \ref{fig:ESR-spectra-of}
and \ref{fig:The-ESR-spectra}). Usually it took about 2 days to accumulate
sufficiently strong ESR lines in ``pure'' neon samples, while the
H atom ESR lines in Samples with 1\% and higher H$_{2}$ admixture
appeared within a few hours.

\begin{figure}
\includegraphics[width=1\columnwidth]{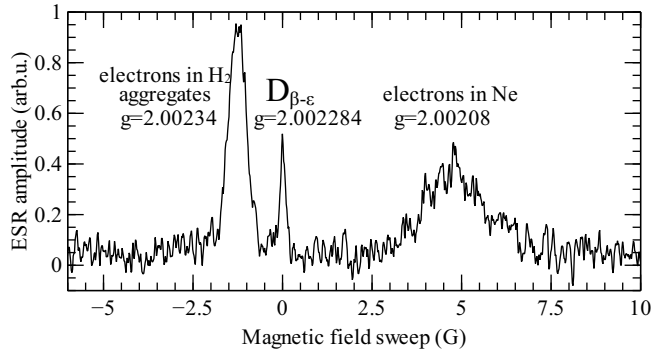}

\protect\caption{Magnified center of the ESR spectrum measured in Sample 1 (a ``pure''
Ne film) after condensing helium film into the sample cell and accumulating
signal of deuterium atoms in the gas phase. The D$_{\beta-\epsilon}$
line of free D atoms in the gas phase ($g$=$2.002284$ \cite{Vanier89})
was used as a magnetic field reference for determining the $g$-factors
of other ESR lines in the figure (Table \ref{tab:spectroscopic-parameters}).
See Page 8 for a detailed discussion. \label{fig:Electrons}}
\end{figure}

In addition to the lines of H atoms, we also observed two singlet
lines close to the center of the ESR spectrum with $g$=2.00208 and
$g$=2.00234 which we attributed to electrons trapped in solid neon
and hydrogen rich regions (Fig.$\,$\ref{fig:Electrons}). Similar
lines with $g$-factors nearly equal to that of free electrons were
observed previously in different matrices of hydrogen isotopes \cite{Collins96,Electrons_QFS}.

The shape of all ESR lines of H atoms in the different Ne:H$_{2}$
films we studied had a somewhat similar structure as shown in Fig.$\,$\ref{fig:The-ESR-spectra}.
The spectra were measured after 1.5-2 days of sample storage at $T$=90$\,$mK.
Fitting of the ESR lines of atomic hydrogen in neon for Sample 3 (Ne:1\%
H$_{2}$) by three Lorentzian lines is presented in Fig.$\,$\ref{fig:ESR-spectra-of}.
In order to make the fitting, we carefully measured the electronic
$g$-factors of the H$_{bc}$ line components using the reference
line of free hydrogen atoms in the gas phase. The values of the hyperfine
constants, $A$, for H atom doublets were measured using the ENDOR
technique as described further. The positions of all three components
of the H$_{ad}$ line as shown in Fig.$\,$\ref{fig:ESR-spectra-of}
were calculated based on the $g$-factors extracted for the H$_{bc}$
line as well as measurement of the hyperfine constant by ENDOR, and
the fitting was carried out having the component amplitudes as the
only adjustable parameters. The values of $A$ measured by ENDOR for
different samples were nearly similar within the uncertainty (a few
hundreds kHz), associated with the width of the observed ENDOR transitions.
Therefore, we only present the spectroscopic parameters for H atom
ESR line components for intermediate Sample 3 (Table \ref{tab:spectroscopic-parameters}).
The width of Components 1 and 2 increased in the samples with a high
(3 and 6\%) H$_{2}$ admixture which might be a result of their partial
merging.

\begin{table*}
\begin{tabular}{|c|c|c|c||c|c||c|}
\hline 
\multicolumn{7}{|c|}{Present work}\tabularnewline
\hline 
 & C1  & C2  & \multicolumn{1}{c|}{C3 } & H in H$_{2}$ & \multicolumn{1}{c|}{$e^{-}$ in Ne } & $e^{-}$ in H$_{2}$ clusters\tabularnewline
\hline 
\hline 
$g_{e}$  & 2.00229(2)  & 2.00223(2)  & \multicolumn{1}{c|}{2.00222(2) } & 2.00229(1) & \multicolumn{1}{c|}{2.00208 } & 2.00234 \tabularnewline
\hline 
$A$ (MHz) & 1417.7(4)  & 1419.0(3)  & \multicolumn{1}{c|}{1426.2(5) } & 1417.40(2) & \multicolumn{1}{c|}{} & \tabularnewline
\hline 
$\Delta A$ (MHz) & -2.7(4) & -1.4(3) & \multicolumn{1}{c|}{5.8(5)} & 3.01(2) & \multicolumn{1}{c|}{} & \tabularnewline
\hline 
$\Delta A/A$ (\%)  & -0.19(3)  & -0.10(2)  & \multicolumn{1}{c|}{+0.40(3) } & -0.21(1) & \multicolumn{1}{c|}{} & \tabularnewline
\hline 
Width (G)  & 0.8  & 1.3  & \multicolumn{1}{c|}{1.8 } & 1.0 & \multicolumn{1}{c|}{2.1 } & 0.6 \tabularnewline
\hline 
\hline 
\multicolumn{7}{|c|}{Previous studies}\tabularnewline
\hline 
 & C1  & C2  & C3  & C1 & C2  & C3\tabularnewline
\hline 
\hline 
$g_{e}$  &  &  &  & 2.00213(8)  & 2.00211(8)  & 2.00207(8) \tabularnewline
\hline 
$A$ (MHz) & 1417.4(2)  & 1418.99(15)  & 1426.11(15)  & 1417.4(2)  & 1418.5(3)  & 1426.56(20)\tabularnewline
\hline 
$\Delta A$ (MHz) & -3.0(2) & -1.42(15) & 5.70(15) & -3.0(2) & -1.9(3) & 6.15(20)\tabularnewline
\hline 
$\Delta A/A$ (\%)  & -0.21(1)  & -0.10(1)  & 0.40(1)  & -0.21(1) & -0.13(2)  & 0.43(1) \tabularnewline
\hline 
Width (G)  &  & 0.09  & 0.15-0.30  & 0.5  & 0.09  & \tabularnewline
\hline 
Ref. & \multicolumn{3}{c||}{\cite{Dmitriev89}} & \multicolumn{2}{c||}{\cite{Zhitnikov87}} & \cite{Foner60}\tabularnewline
\hline 
\end{tabular}

\protect\caption{The main spectroscopic parameters measured for H atoms and electrons
in Sample 3 (Ne:1\% H$_{2}$) and Sample 7 (pure H$_{2}$) of the
present work and those for H atoms in solid neon observed in previous
studies \cite{Dmitriev89,Zhitnikov87,Foner60}. \label{tab:spectroscopic-parameters} }
\end{table*}

\subsection{ENDOR}

The interaction of the guest H atoms with the host particles is expected
to slightly change both their electronic $g$-factors and the hyperfine
constants $A$ compared to those of free atoms in the gas phase. The
sign of a hyperfine constant change characterizes the lattice sites
the H atoms occupy. It is negative for spacious substitutional positions
and positive for more cramped interstitial sites \cite{Adrian60}.
We associate the difference in shapes of the high- and low-field hydrogen
lines with the presence of three different positions of H atoms in
the matrix with different hyperfine constants and electronic $g$-factors.
Atoms in each position produce components in the ESR spectrum with
different separations between the lines of the H doublet defined by
the value of $A$, and different positions of the doublet center defined
by the corresponding $g$-factors. It turns out that the components
with larger $A$ (Component 3 in Fig.$\,$\ref{fig:ESR-spectra-of})
have a smaller $g$-factor and the center of the doublet is shifted
to the right, towards a higher sweep field. Therefore, these contributions
nearly compensate each other for the H$_{ad}$ line and the different
components appear almost merged, while they diverge for the H$_{bc}$
line, where both the hyperfine constant and $g$-factor change increase
the line shifts.

\begin{figure}
\includegraphics[width=1\columnwidth]{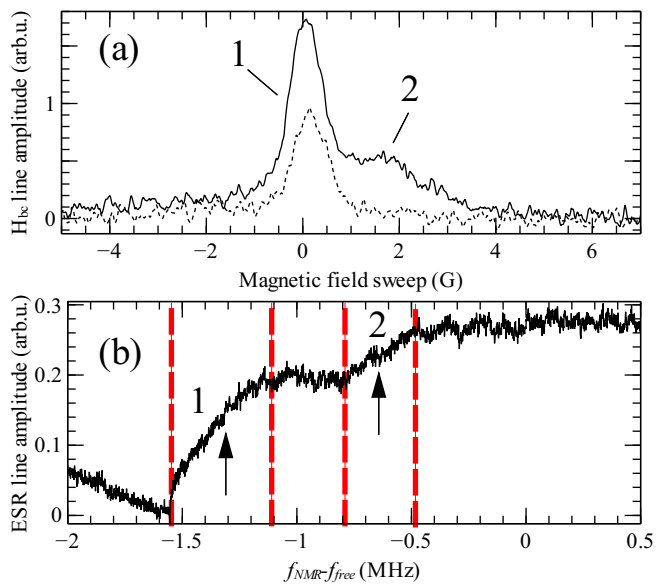}\protect\caption{H$_{bc}$ line before (dashed) and after (solid) measuring ENDOR spectrum
(a). ENDOR spectrum showing transitions recovering the first (C1)
and second (C2) component of the H$_{bc}$ line measured in Sample
6 (Ne/H$_{2}$) (b). Note that only Components 1 and 2 in (a) increased.
The ENDOR transitions margins are designated by red dashed lines.
The resonance positions for each transition are marked by arrows.
\label{fig:ENDOR_1_2}}
\end{figure}

\begin{figure}
\includegraphics{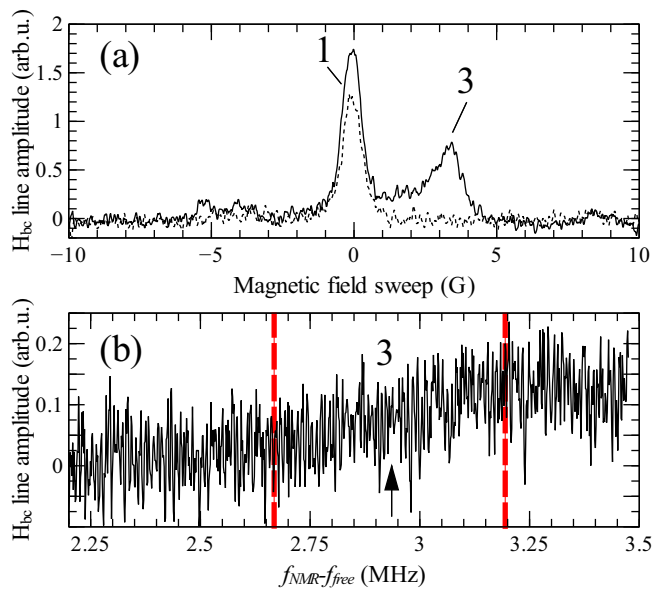} \protect\caption{H$_{bc}$ line before (dashed) and after (solid) measuring ENDOR spectrum
(a). ENDOR spectrum recovering the third (C3) H$_{bc}$ line component
measured in Sample 6 (Ne/H$_{2}$) (b). Note that only Component 3
in (a) increased. The small increase of Component 1 is due to the
relaxation process. The transition margins are designated by red dashed
lines The resonance position is marked by arrow. \label{fig:ENDOR_3} }
\end{figure}

We performed accurate measurements of the hyperfine constant change
for all three components of the H ESR lines using the ENDOR technique.
The method is based on determination of the NMR transition frequency
indirectly by its influence on the ESR line amplitude when the rf
excitation frequency exactly matches the $a-b$ transition frequency
(see Fig.$\,$\ref{fig:LevelDiag1}). Prior to measuring ENDOR transitions,
we created a Dynamic Nuclear Polarization (DNP) using the Overhauser
effect \cite{DNPPRL14} in order to decrease the H$_{b}$ level population
and enhance the ENDOR signal. The procedure is based on saturation
of the allowed ESR, $b-c$, transition with a subsequent cross-relaxation
via the forbidden $c-a$ transition (Fig.$\,$\ref{fig:LevelDiag1}).
Using this DNP method we were able to transfer all three components
of the H$_{bc}$ line to the H$_{ad}$ line separately or transfer
the whole line using 15$\,$MHz FM modulation of the ESR spectrometer
frequency. Then, stopping the magnetic field sweep at the position
of one of the components (C1, C2, or C3) of the H$_{bc}$ ESR line,
we performed slow scans of the rf frequency near the resonance value
for the NMR $a-b$ transition (close to $f_{ab}\approx$ 910$\,$MHz
for free atoms). By recording the ESR signal amplitudes as a function
of the applied RF frequency we retrieved the ENDOR spectra for each
component of the H$_{bc}$ line. The ENDOR spectra are presented in
Figs. \ref{fig:ENDOR_1_2} and \ref{fig:ENDOR_3}. For a slow enough
sweep rate, the $a-b$ transition was saturated during the rf sweep
and the populations of the H$_{b}$ and H$_{a}$ levels were equalized.
As a result, the corresponding component re-appeared in the ESR H$_{bc}$
line spectrum since the H$_{b}$ level now became well populated (see
Figs \ref{fig:ENDOR_1_2}a and \ref{fig:ENDOR_3}a).

Pumping the H$_{bc}$ line without modulation burned a 0.1$\,$G wide
hole in the H$_{bc}$ line which was reproducible for all three components.
The hole formation corresponds to saturation of an individual group
of spins and its width provides a contribution from homogeneous broadening
which is determined by the dipolar interaction between the electron
spins of atoms.

We observed three different ENDOR transitions which recovered all
three components of the H$_{bc}$ line. The hyperfine constant can
be determined from the measured transition frequency according to
the formula \cite{Ahokas10} 
\begin{equation}
A\simeq2f_{NMR}-\frac{\gamma_{H}B}{\pi}\label{eq:ENDOR}
\end{equation}
where $\gamma_{H}$ is the proton gyromagnetic ratio and $B$ is the
static magnetic field. We found that the ESR lines were easily saturated
which made the registration of ENDOR spectra rather difficult. We
present the ENDOR spectra measured for Sample 6 (Ne/H$_{2}$) where
they appeared to be unambiguous and clear. The ENDOR transition widths
are solely defined by the spread of hyperfine constants $A$ for each
component. Therefore, we constrain the uncertainty of determining
$A$ for each component by the width of a corresponding ENDOR transition.
The spectroscopic parameters for Samples 1-6 coincided within the
experimental uncertainty. Therefore, we present in Table \ref{tab:spectroscopic-parameters}
only the values of $A$ and $g$ for an intermediate Sample 3, Ne:1\%H$_{2}$.

The first two ENDOR transitions are shown in Fig.$\,$\ref{fig:ENDOR_1_2}.
Transitions 1 and 2 recovered the left (C1) and the central components
(C2) of the H$_{bc}$ line, respectively. The first transition corresponds
to a hyperfine constant change of -2.6(4)$\,$MHz. The second transition
corresponds to a smaller hyperfine constant shift, $\Delta A$=$-$1.4(3)$\,$MHz.
Both transitions, 1 and 2, are characterized by negative hyperfine
constant changes and can be attributed to the substitutional positions
in the matrix \cite{Adrian60}. The second transition appeared only
in Ne:H$_{2}$ mixtures (Samples 1-6), while in pure H$_{2}$ (Sample
7) only a transition similar to transition 1 was observed. 

The third ENDOR transition shown in Fig.$\,$\ref{fig:ENDOR_3} recovered
the right H$_{bc}$ line component (C3) and corresponded to a hyperfine
constant change of $+5.8(5)\,$MHz. This line was observed previously
\cite{Foner60,Knight98,Dmitriev89} and was associated with H atoms
in somewhat cramped substitutional sites of solid Ne. The spectroscopic
parameters for pure H$_{2}$ (Sample 7) are presented in Table \ref{tab:spectroscopic-parameters}
for comparison.

\begin{figure}
\includegraphics[width=1\columnwidth]{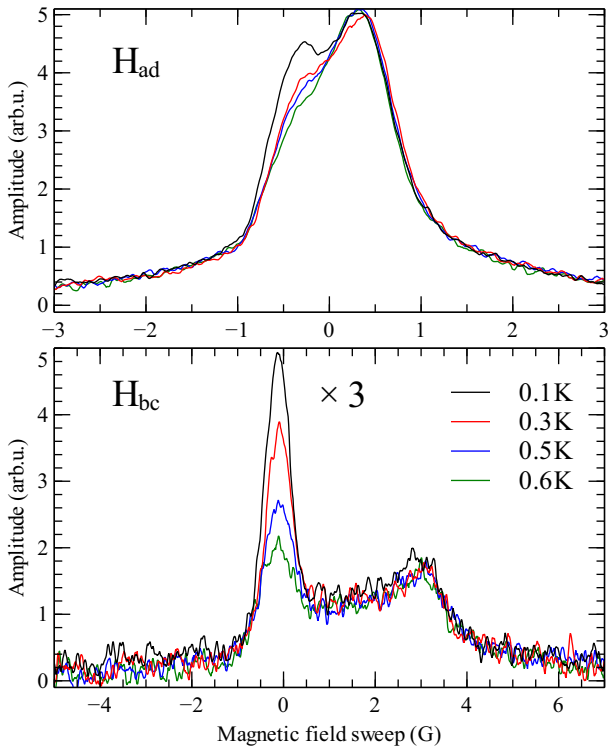}\protect\caption{Effect of step-wise raising temperature in the range from 0.1 to 0.6$\,$K
on ESR lines of H atoms in Sample 3 (Ne:1\%H$_{2}$). Note that only
the left peak corresponding to Component 1 in both lines changes.
Note that the H$_{bc}$ line amplitude is multiplied by a factor of
3.\label{fig:Effect-of-raising-temperature}}
\end{figure}

\subsection{Effect of temperature and helium film on ESR spectra}

After storing the samples for 1-3 days to accumulate H atoms produced
by tritium decay, we condensed small amounts of helium into the sample
cell sufficient to form a superfluid film covering the sample cell
walls. The film is required for accumulation and storage of the H
and D atoms in the gas phase to be used as the magnetic field markers.
We found that the admission of helium had an immediate and remarkable
influence on the ESR spectra of all as-deposited samples as described
below. Condensing the He film into the sample cell was naturally accompanied
by a rise of temperature from 0.1 to 0.6-0.7$\,$K, resulted from
the refluxing helium vapor. In order to separate out the effect of
sample heating, prior to condensing He, we performed a separate study
of the effect of temperature on the properties of the samples in the
temperature range from 0.1$\,$K to 0.6$\,$K.

We found that raising the temperature leads to an abrupt decrease
of the component C1 of the H ESR spectrum. This effect was somewhat
different for the samples with different concentrations of H$_{2}$.
The strongest drop of the signal (by $\sim70\%$) was observed for
Sample 3 with 1$\%$ of H$_{2}$ concentration. The effect was very
weak for the ``pure'' neon Sample 1, whereas $\simeq$25\% of atoms
disappeared in Samples 2 (Ne:0.2\% H$_{2}$), 4 (Ne:3\% H$_{2}$)
and 5 (Ne:6\% H$_{2}$).

\begin{figure}
\includegraphics{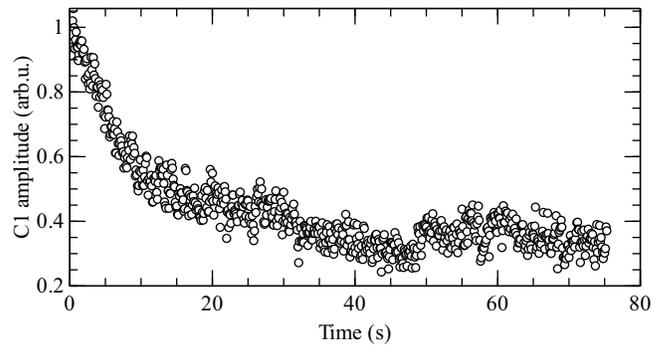}\protect\caption{Time evolution of the C1 amplitude in Sample 3 (Ne:1\%H$_{2}$) after
an instantaneous rise of temperature from 0.1 to 0.6$\,$K. \label{fig:C1_heat_pulse}}
\end{figure}

To get further insight into this phenomenon, we studied the behavior
of the C1 intensity in response to a series of step-like increases
of temperature by 100-200$\,$mK. The temperature was raised and stabilized
with a characteristic time of $\approx30\,$sec. ESR spectra were
recorded with the same time interval. After each step we observed
a drop of the C1 intensity which was detected even for the first recorded
ESR spectrum, as shown in Fig.$\,$\ref{fig:Effect-of-raising-temperature}.
The ESR line intensity did not change further until the next step.
The total decrease of the C1 intensity after heating to 0.6 K was
$\approx70\%$. In order to better resolve the dynamics of this effect
we applied a heat pulse to the sample cell while standing at the C1
peak maximum and measuring the C1 component amplitude as a function
of time after the pulse. We adjusted the energy of the pulse to heat
the cell from 0.1 to 0.6$\,$K within several seconds. The evolution
of the C1 amplitude after the heat pulse is presented in Fig.$\,$\ref{fig:C1_heat_pulse}.
The characteristic decay time extracted from such a measurement is
$\simeq$10$\,$s. In fact, this decay time is an upper limit estimate
of the actual decrease time, since the thermal response time of the
sample cell to the heating pulse is of about the same order.

One can see in Fig.$\,$\ref{fig:Effect-of-raising-temperature} that
the heating steps did not influence the other components of the H
ESR spectrum. We measured the eventual recovery of Component 1 in
Sample 3 after destruction by heating to 0.6 K and found that it recovered
with the same speed as for its growth before destruction. This leads
us to the conclusion that the observed rapid decay is not related
to spin-relaxation effects, but is caused by the recombination of
atoms.

It should be emphasized that the temperatures 0.3-0.6$\,$K are extremely
low, and in our previous work, we were only able to detect a tiny
decrease of the ESR signal due to recombination of H atoms in H$_{2}$
films on the time scale of several days \cite{Ahokas10}. Recombination
of H atoms on the time scale of seconds provides evidence of an extremely
high recombination rate. For the estimate of the rate constant we
need to know the local density of the H atoms in the nanocluster of
H$_{2}$. Here the main uncertainty comes from the unknown volume
of these clusters and distribution of atoms in them. Therefore, the
most reasonable estimate of the density can be done from the homogeneous
contribution to the C1 line width caused by the dipole-dipole interactions
between atoms, which is proportional to the H density $\Delta H_{hom}[G]\approx0.85\cdot10^{-19}n$
{[}cm$^{-3}${]} \citep{Ahokas10}. This broadening was measured by
burning a hole in the ESR line, which resulted in the hole width of
$\approx0.1$ $\,$G. Since there can be other mechanisms of the inhomogeneous
broadening this provides an upper limit estimate of the H density
$n\leq1.3\cdot10^{18}$$\,$cm$^{-3}$. Then, assuming the second
order recombination process $dn/dt=-2k_{r}n^{2}$ and using an upper
limit estimate for the half-decay time $\tau\leq10$ s, we obtain
a lower limit estimate for the recombination rate constant $k_{r}\geq5\cdot10^{-20}$ $\,$cm$^{3}$s$^{-1}$.
This has to be compared with the previously measured recombination
rate constant of H atoms in the solid H$_{2}$ at the same temperature
$k_{r}\sim(2-10)\cdot10^{-25}$$\,$cm$^{3}$s$^{-1}$ \cite{HinH2_06,Ahokas10}.

In order to study the influence of helium on the sample properties,
we stabilized the cell temperature at 0.6$\,$K and He gas was slowly
condensed in 2-$\mu$mole portions. Condensing helium led to a nearly
complete sudden recombination of H atoms corresponding to Component
1 in Samples 1-3 with low concentration of hydrogen. The decrease
of the C1 intensity was also observed for Samples 4-5, but the line
did not completely disappear in this case. This change of the C1 intensity
was irreversible in the sense that no further growth was observed
even after pumping out helium from the sample cell and cooling it
to 0.1$\,$K.

For Samples 1-3 in addition to the decrease of C1 in the H spectrum
we found that a new ESR line ($g$=2.00234, see Fig.$\,$\ref{fig:Electrons})
started growing in the spectrum center immediately after admission
of helium. This line growth changed to a decrease after pumping out
helium from the sample cell, and eventually it vanished in the noise.
We conclude that for the low concentration samples 1-3, the presence
of helium is necessary to observe the narrow electron line. The situation
was different for the samples with a 3\% and higher H$_{2}$ admixtures,
the same electron line ($g$=2.00234) appeared immediately after deposition
of the samples, and condensing He did not affect its width or amplitude.

\begin{figure}
\includegraphics[width=1\columnwidth]{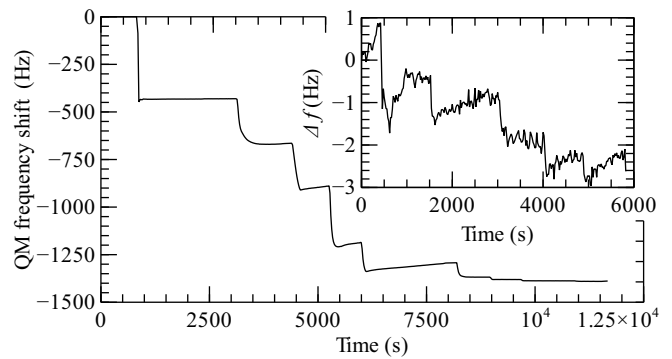}\protect\caption{The quartz microbalance frequency shift after condensing helium while
storing the as-deposited (main plot) and annealed ``pure'' Ne Sample
6 (inset). Each step-like quartz-microbalance frequency shift corresponds
to a condensation of 2$\mu$mol of He into the sample cell. \label{fig:The-quartz-microbalance}}
\end{figure}

Condensing helium into the sample cell with neon films resulted in
a significant shift of the quartz microbalance frequency, $\sim$1500$\,$Hz,
which was reproducible for all as-deposited samples. It should be
emphasized that only non-superfluid He layers deposited on the quartz
microbalance can be detected while the superfluid fraction decouples
from the QM oscillations and does not contribute to the frequency
shift. Thus the saturated superfluid film condensed into the empty
cell leads to the QM shift of several Hz corresponding to several
normal layers of helium. The QM frequency shift upon condensing He
into the cell for ``pure'' Ne as-deposited Sample 6 (Fig.$\,$\ref{fig:The-quartz-microbalance})
is the largest for the first 2-$\mu$mole He portion ($\sim$450$\,$Hz)
and it gradually decreases for the following ones. Condensing the
last 2-$\mu$mol He portions results only in a minor frequency shift
(<1$\,$Hz) which allows us to conclude that the process is saturated
and all pores are filled by helium. The observed QM frequency shift
of 1500$\,$Hz is equivalent to adsorption of $\sim$1100 non-superfluid
He monolayers on the Ne film with a thickness of 11000 monolayers.

\begin{figure}
\includegraphics[width=1\columnwidth]{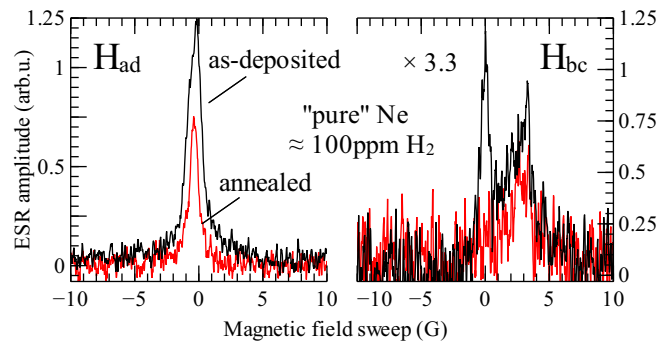}\protect\caption{H atom ESR lines measured in ``pure'' Ne Sample 6 before (black)
and after sample annealing (red). Note that the H$_{bc}$ line amplitude
is multiplied by a factor of 3.3.\label{fig:ESR-lines-annealing}}
\end{figure}

A large shift of the quartz-microbalance oscillation frequency after
condensing helium allows us to suggest a large porosity of the as-deposited
neon films. In order to eliminate this and make crystals more regular
we performed an annealing procedure. The cell temperature was increased
to $T$=7-10$\,$K for 1.5 hours. The annealing temperatures were
chosen to be higher than 1/4 of the Ne melting temperature \cite{Nepijo05}.
This was sufficient for a partial sublimation of the neon film $\sim10-15\%$
to occur. After annealing, the sample cell was cooled to 0.6$\,$K
and the experimental procedure of condensing helium was carried out
again. We found that the annealing procedure worked well for samples
with the smallest concentration of H$_{2}$. The QM frequency shift
due to admission of He for the annealed (\textquotedbl{}pure\textquotedbl{}
neon) Sample 6 was only $\simeq$5$\,$Hz (inset in Fig.$\,$\ref{fig:The-quartz-microbalance}).
We did not observe any film erosion effects caused by electron bombardment
\cite{Schou86}. Condensing helium into the cell right after annealing
and after storing the annealed samples for 2 days led to the same
QM frequency shift (about 5$\,$Hz).

For the samples with larger concentrations of hydrogen, the above
described annealing procedure did not work well. The helium-related
QM shift decreased by the factor of 3, but still remained at the level
of 400-500 Hz, allowing us to assume that substantial porosity remained
and could not be removed by the annealing procedure used.

The influence of annealing on the ESR spectrum was different for samples
with different concentrations. For the ``pure'' Ne Sample 6 (Fig.$\,$\ref{fig:ESR-lines-annealing})
no helium was added, and the ESR spectrum contained components 1-3
before annealing. The H ESR lines became narrower after annealing
and contained only Component 3. Annealing samples with higher concentrations
led to further decrease of the C1 component. We recall that part of
it disappeared in recombination after heating and helium admission.
All components remained in the spectrum and became more narrow and
better resolved.

\subsection{Efficiency of H atom accumulation in solid Ne samples}

\begin{figure}
\includegraphics[width=1\columnwidth]{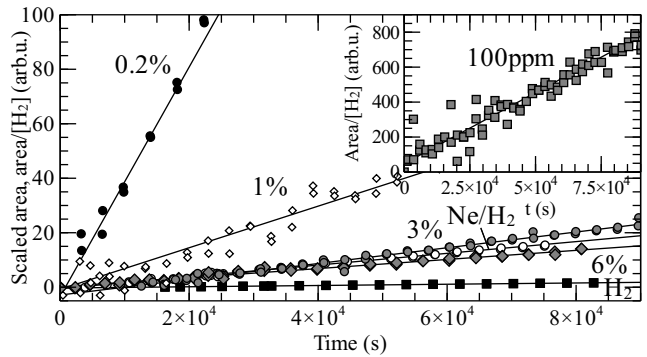}\protect\caption{Accumulation rates of H atoms in the samples described in this work.
The Y-axis units for each plot are scaled to the estimated number
of H$_{2}$ molecules in each sample, $i.e.$ the ESR line integral/{[}H$_{2}${]}.
The H$_{2}$ admixtures in Ne are labeled near each data set. Ne/H$_{2}$
should be understood as a 160$\,$nm H$_{2}$ film deposited on top
of the annealed neon film (Sample 6). The H$_{2}$ admixture in a
``pure'' neon sample was considered as 10$^{-4}$ (100ppm).\label{fig:Accumulation-rates-of}}
\end{figure}

All as-prepared samples were stored for several days in order to estimate
the accumulation rates of H atoms. The flux of electrons resulting
from decay of tritium in the sample cell walls was essentially the
same for all samples, and the growth rate of the signals represent
the efficiency of the dissociation process. We found that the growth
rate of the H atom signal in ``pure'' neon films containing 100$\,$ppm
H$_{2}$ was only 20 times smaller than that in pure H$_{2}$ of the
same thickness. Even more striking was the enhancement of the accumulation
rate, $d[\mbox{H}]/dt$, which was observed in Samples 3 and 4 where
it matched that in pure hydrogen. In order to compare the efficiency
of dissociation we scaled the growth of H atom ESR signals to an estimated
number of H$_{2}$ molecules in each sample. The ratio of H atom ESR
line integrals to H$_{2}$ content in the neon-hydrogen film is displayed
in Fig.$\,$\ref{fig:Accumulation-rates-of}. We found that the H
atom accumulation efficiency in Ne:H$_{2}$ solid mixtures is much
larger in comparison to that for pure H$_{2}$ samples and it has
a tendency to increase when the H$_{2}$ admixture in solid Ne decreases.

We considered two possible explanations for a smaller efficiency of
H atom accumulation in solid hydrogen compared to that in solid neon-hydrogen
mixtures. The H atoms generated by $\beta$-particles in solid H$_{2}$
may recombine during thermalization, and the resulting fast recombination
might lead to decreasing of their yield. Previously we found out that
every $\beta$-particle generates about 50 H atoms in a solid H$_{2}$
film \cite{Tritium_paper} which is more than an order of magnitude
smaller as compared to that observed in the gas-phase \cite{Collins92}.
Although the matrix may provide pathways for an excited molecule to
relax back to the ground state without being dissociated, it might
be considered that a significant number of H atoms in pure H$_{2}$
instantaneously recombine back, thus reducing the H atom yield. The
second possibility we considered is an enhancement of the accumulation
rate in the neon samples due to the action of secondary electrons
which significantly amplify production of H atoms. To prove the second
mechanism we performed an experiment in which we created a layered
Ne/H$_{2}$ sample where a 160$\,$nm H$_{2}$ layer was deposited
on top of the annealed neon film (Sample 6). We found that the accumulation
efficiency of H atoms in the H$_{2}$ layer deposited on top of a
Ne film was nearly 9 times higher as compared to pure H$_{2}$ films.
The penetration depth of electrons with energy of 5.7$\,$keV into
solid H$_{2}$, $\simeq$3.5$\,$$\mu$m \cite{Schou78}, exceeds
the thickness of pure hydrogen films we studied which should insure
a rather uniform generation of H atoms both in thick, $2.5\,\mu$m
solid Ne and 160$\,$nm H$_{2}$ films. This led us to the conclusion
that the enhancement of H atom accumulation efficiency in solid neon-hydrogen
films is indeed caused by a large yield of secondary electrons. The
enhancement of accumulation efficiency at very low H$_{2}$ concentrations
in solid Ne, on the other hand, provides evidence for a suppression
of atomic hydrogen recombination under these conditions.

\section{Discussion}

In this work we report observations and measurements of the main spectroscopic
parameters of three doublets of H atoms in the ESR spectra. The values
of these parameters are in good agreement with previous studies \cite{Foner60,Zhitnikov87,Dmitriev89}
(see Table \ref{tab:spectroscopic-parameters}). Dmitriev et al. \cite{Dmitriev89}
observed the H atom ESR doublets characterized by both positive and
negative hyperfine constant shifts in their Ne:H$_{2}$ samples created
by co-deposition of the rf-discharge products at small H$_{2}$ concentrations
(10-100$\,$ppm) at $T$=4.2$\,$K. The spectroscopic parameters obtained
in their work appeared to be similar to Components 3 and 2, respectively,
in our experiments. The spectroscopic parameters of H atoms in solid
Ne described in refs. \cite{Foner60,Zhitnikov87} are presented in
Table \ref{tab:spectroscopic-parameters}. They are nearly same as
reported here.

The ENDOR transitions provide information on the change of the hyperfine
constant which is most sensitive to the environment of impurity atoms
and allows characterization of their position in the solid neon-hydrogen
matrix. The characterizations which we consider below are in good
agreement with conclusions made in the previous studies \cite{Foner60,Zhitnikov87,Dmitriev89}.

Component 1 has a negative hyperfine constant change, $\Delta A=-2.6$(4)$\,$MHz,
$A$=1417.8(4)$\,$MHz. This is somewhat smaller than the hyperfine
constant change for H atoms in a pure H$_{2}$ environment $\Delta A=-3.0(2)$$\,$MHz.
We suggest that the small reduction of $\Delta A$ may be caused by
a presence of both H$_{2}$ molecules and Ne atoms in the closest
neighborhood of H atoms. We assign the C1 component to the hydrogen
atoms trapped in pure H$_{2}$ regions, clusters or micro-crystals
embedded in the neon matrix.

The polarizability of Ne atoms is even smaller than that of H$_{2}$
molecules which should result in a smaller hyperfine constant change.
Therefore, we suggest that Component 2, characterized by $\Delta A$=-1.4(3)$\,$MHz,
$A$=1419.0(3)$\,$MHz, is related to H atoms in the substitutional
sites of the Ne lattice.

The third component C3 is characterized by a clearly positive hyperfine
constant change, $\Delta A$=5.8(5)$\,$MHz, $A$=1426.2(5)$\,$MHz.
Even though the shift is positive, its absolute value is too small
for the pure interstitial position. It might be expected that the
extremely small polarizability of Ne atoms will lead to a dominance
of the Pauli repulsion even at long distances. Therefore, we attribute
the C3 component to H atoms which were initially stabilized in octahedral
interstitial voids of the Ne lattice but then relaxed to a position
which is somewhat intermediate between a substitutional and a more
cramped octahedral interstitial one, as was suggested theoretically
by Kiljunen et al. \cite{Kiljunen99}.

From the previous studies it is known that H atoms can occupy different
lattice sites depending on the substrate temperature and deposition
rate\@. Vaskonen et al. \cite{Vaskonen99} observed two ESR line
doublets of H atoms in solid Ar and Kr and associated them with H
atoms in substitutional and octahedral interstitial sites, respectively.
The doublet assigned to the substitutional sites of H atoms in solid
Ar and Kr appeared stronger for colder substrates and higher deposition
rates, conditions which favor formation of vacancies. Annealing the
sample decreases the number of vacancies and more atoms become trapped
in the octahedral interstitial sites. A similar result was observed
in our work where only Component 3 remained after annealing a ``pure''
Ne sample.

The new and unexpected observation is the strong influence of temperature
and the superfluid helium film on the properties of atoms trapped
in clusters of pure H$_{2}$ (C1 component), which we first summarize
and then try to explain below.

Raising the temperature in the range 0.1-0.6$\,$K$\,$leads to a
rapid recombination of a fraction of H atoms corresponding to the
C1 component. The rest of these atoms remain stable until a further
temperature increase is performed. The effect is strongest for the
1$\%$ concentration, where 2/3 of the C1 component is destroyed by
raising the temperature from 0.1 to 0.6$\,$K. The recombination is
very fast and occurs on the time scale comparable with the thermal
response of the sample cell to the heating pulse. We evaluated the
lower limit for the recombination rate constant $k_{r}\geq5\cdot10^{-20}$$\,$cm$^{3}$s$^{-1}$,
to be nearly five orders of magnitude larger than that measured previously
in pure H$_{2}$ films \cite{Ahokas10}. Lowering the temperature
back to 0.1 K leads to a recovery of the C1 component with the same
accumulation rate observed during initial H atom accumulation in this
sample. Admission of helium film completely destroys the C1 component
in the samples with small concentrations of hydrogen ($\leq1\%$).
The destruction is irreversible: pumping out helium and cooling back
to 0.1$\,$K does not lead to the growth of the C1 component. For
higher concentrations of H$_{2}$ in Ne the destruction by admission
of helium is only partial, with most of the C1 atoms surviving. Annealing
of the samples leads to the complete disappearance of the C1 component
at small concentrations ( $\leq1\%$ samples 1-3), while for larger
concentration (samples 4,5) part of the C1 line survives the annealing
procedure.

The very fast changes which occur with the hydrogen atoms trapped
in the pure H$_{2}$ regions upon heating, motivates us to consider
a possible phase transition as the main effect behind the observed
behavior.

First we consider possible phases of hydrogen in neon. Gal'tsov et
al. \cite{Gal'tsov04} showed that solid Ne:H$_{2}$ mixtures with
2-12\% H$_{2}$ admixtures are characterized by a co-existence of
the $fcc$ and $hcp$ phases of solid neon with nearly identical lattice
parameters. The latter phase is rich in H$_{2}$ and appears stable
up to about 16$\,$K, while the former may accommodate not more than
$\simeq$2\% H$_{2}$ without being driven out of equilibrium and
transforming to the $hcp$ phase which is able to store larger amounts
of H$_{2}$. It might be suggested that the C1 component we observe
corresponds to H atoms in the regions rich in molecular hydrogen corresponding
to a highly unstable $fcc$ lattice. This phase might be stabilized
by defects whereas raising the temperature or condensing helium might
trigger the phase transition. One might expect that the possible transition
of the unstable hydrogen-rich $fcc$ neon lattice to the more stable
$hcp$ phase should be irreversible and appear only once. However,
the C1 component recovered on a time scale of hours if the cell temperature
was returned to 0.1$\,$K and atomic recombination could be triggered
again by raising the temperature to 0.3-0.6$\,$K. Such fast recombination
which we observed cannot appear in any solid matrix, irrespective
of its type. On this basis we concluded that the above described behavior
is unlikely to be related to the $fcc$ - $hcp$ phase transition
in Ne:H$_{2}$ solid mixtures.

The other possibility for this phase transition in hydrogen is the
solid-to-liquid transition in H$_{2}$ clusters. Clearly, if the H
atoms were in the liquid, they would have a very high mobility, which
could explain the fast recombination of C1 component. It is known
that for molecular hydrogen in a restricted geometry, the solidification
temperature may be substantially reduced. This is confirmed in experiments
in porous media \cite{Tell83,Molz93,Sokol96}. The liquid behavior
of small hydrogen clusters ($\sim10^{4}$ molecules) in helium nanodroplets
has been observed in the experiments of Kuyanov-Prozument and Vilesov
\cite{Kuyanov08} at temperatures $\sim2$$\,$K. Theoretical calculations
predict liquid and even superfluid behavior for small ($\leq30$ molecules)
clusters of H$_{2}$ at temperatures below 1$\,$K \cite{Sindzingre91,Kwon02,Mezzacapo06},
while for large size of clusters no such prediction exists. The superfluid
behavior had a possible experimental confirmation in the experiments
of Grebenev \textit{et. al.} \cite{Grebenev00}, but has been under
discussion by other authors \cite{Callegari01,Omiyinka14}.

In our samples the phases of pure H$_{2}$ are definitely formed,
and are associated with the C1 component of the H atom ESR line. We
may assume that these phases form clusters of H$_{2}$ molecules of
different sizes with a distribution around the mean value. The average
size should depend on the concentration of hydrogen in the condensed
neon-hydrogen mixture. Once we observe strong porosity of as-deposited
films, it is natural to suggest that all H$_{2}$ clusters or at least
part of them are located in the pores of the Ne matrix. With these
assumptions we arrive at a system with small hydrogen clusters in
a confined geometry where a solid-to-liquid transition may occur at
very low temperatures. The H$_{2}$ clusters accumulate atomic hydrogen
after dissociation enhanced by secondary electrons. We note that the
estimated concentration of atoms $n\leq10^{18}$ cm$^{-3}$ corresponds
to one atom per 10$^{4}$ molecules or less, and therefore the size
of the clusters accommodating several H atoms should be substantially
larger than that. The solid-liquid transition temperature should depend
on the size of the cluster. Therefore, by raising the sample temperature,
we trigger the solid-liquid transition in a fraction of them having
the proper size. The atoms inside these clusters acquire a high mobility
inside the liquid and rapidly recombine, while the others, remaining
solid, keep their atoms stabilized in the H$_{2}$ matrix. A subsequent
heating step involves another fraction of the clusters, and so on.
Cooling back to 0.1$\,$K transfers them back to the solid state,
and accumulation of atoms starts with the same rate. For high concentrations
of hydrogen the average size is shifted towards very large clusters,
which remain solid in the whole temperature range of experiment. That
is why for the samples 4-5 the fraction of recombined atoms upon heating
is smaller. It seems that the sample with 1$\%$ concentration of
H$_{2}$ has the average size of clusters which best matches the solid-to-liquid
transition in the studied temperature range 0.1-0.6$\,$K, which explains
why the effect is strongest for this sample.

When we condense helium, it diffuses into the pores and voids of the
sample, and is collected at the boundary between the H$_{2}$ clusters
and the neon crystal. This changes the boundary conditions, and should
obviously change the solid-to-liquid transition temperature. It may
trigger this transition for a fraction of the clusters without any
extra heating. It is also known that pumping out helium will never
remove it all from the system. One or two monolayers will remain adsorbed
on the surface, and the first one may be even in the solid state.
Therefore, once condensed, helium makes a permanent change of the
state for part of the clusters. For small concentrations of H$_{2}$
the average cluster size is such that most of them remain liquid below
1 K and no hydrogen atoms may be accumulated in them. Another feature
of helium which may play important role at the boundary of the clusters
is that it effectively slows down electrons \cite{Dmitriev08,Dmitriev09},
which may enhance their trapping inside the clusters. This is indeed
observed in our experiments by the growth of the narrow component
of the singlet line associated with electrons trapped inside H$_{2}$.
For larger H$_{2}$-Ne concentrations we get a larger fraction of
very large clusters which always remain solid and contain H atoms
with the C1 component. Annealing seems to reduce the number of C1
atoms in the large clusters because of recombination, but probably
does not influence the state of the large clusters. Helium atoms may
still reside at the boundaries between the clusters and the neon lattice,
which have fairly large areas for large H$_{2}$ concentrations. This
explains why relatively large amounts of helium may be still absorbed
by the Samples 3-4 even after annealing.

Another new observation is the enhancement of the accumulation rate
of H atoms in neon samples which appeared to be nearly two orders
of magnitude larger compared to that in pure H$_{2}$ films of the
same thickness. The accumulation rate enhancement was also observed
for H atoms in a H$_{2}$ layer deposited on top of the annealed Ne
film\@. We suggest that this is related to secondary electrons generated
in solid Ne by $\beta$-particles released in the tritium decay. This
agrees qualitatively with the yield of secondary electrons in solid
neon exposed to up to 3$\,$keV primary electrons, $\simeq$70, reported
by Bozhko et al. \cite{Bozhko13}. The accumulation efficiency in
dilute Ne:H$_{2}$ mixtures decreased at higher H$_{2}$ concentrations
which might be related to a higher recombination rate of H atoms in
large clusters of pure H$_{2}$. Similar behavior was observed for
photolysis of HBr in solid Ar, where the HBr dissociation efficiency
estimated from the atomic hydrogen yield increased from 18\% for a
1:500 matrix to 100\% in a 1:8000 matrix \cite{Eloranta99b}.

\section{Conclusions}

We have reported on the ESR study of H atoms stabilized in a solid
neon matrix carried out in a high magnetic field of 4.6$\,$T and
at temperatures below 1$\,$K. The H atoms were generated $in$ $situ$
by electrons released during decay of tritium trapped in the walls
of our sample cell. The ESR lines of H atoms had a complex structure
which was associated with their three different locations: two in
the substitutional positions of neon and the third one in clusters
of pure H$_{2}$.

It was also found that the accumulation of H atoms in Ne:H$_{2}$
solid mixtures is greatly enhanced by the secondary electrons released
from neon atoms upon their bombardment by $\beta-$particles generated
in tritium decay. The accumulation efficiency is even more enhanced
for samples with smaller H$_{2}$ admixtures in solid Ne where H atom
recombination appears to be less efficient compared to samples with
a higher H$_{2}$ abundance. This result may have important applications
for creation of high densities of free radicals in solid matrices

We observed peculiar behavior of the H atoms trapped inside pure molecular
hydrogen clusters located in pores of the neon matrix. Heating in
the temperature range 0.1-0.6$\,$K triggered abrupt recombination
of trapped hydrogen atoms. Such rapid recombination cannot occur in
solids, where the recombination process is limited by the slow diffusion
of atoms. One of the possible explanations suggests a solid-to-liquid
transition for H$_{2}$ clusters of a certain sizes. This may be one
more piece of evidence for the elusive liquid state of H$_{2}$ in
a restricted geometry at ultra-low temperatures. A possible search
for a superfluidity in these clusters could be an intriguing continuation
of this research work.
\begin{acknowledgments}
We acknowledge funding from the Wihuri Foundation and the Academy
of Finland grants No. 258074, 260531 and 268745. This work is also
supported by US NSF grant No DMR 1707565. S.S. thanks UTUGS and the
Turku University Foundation for support. 
\end{acknowledgments}

 \bibliographystyle{apsrev4-1}
\bibliography{HinH2_2015_papers}

\end{document}